\documentclass[12pt]{article}
\usepackage{amsmath,epsfig,graphicx,amssymb}
\usepackage[margin=0.7in]{geometry}
\newcommand{\beq}{\begin{equation}}
\newcommand{\eeq}{\end{equation}}
\newcommand{\ben}{\begin{eqnarray}}
\newcommand{\een}{\end{eqnarray}}
\newcommand{\bes}{\begin{subequations}}
\newcommand{\ees}{\end{subequations}}
\newcommand{\bFig}{\begin{figure}}
\newcommand{\eFig}{\end{figure}}

\date{}
\begin{document}

\title{Unification of Gravity and Electromagnetism I:\\ Mach's Principle and Cosmology}
\author{Partha Ghose\footnote{partha.ghose@gmail.com} \\
The National Academy of Sciences, India,\\ 5 Lajpatrai Road, Allahabad 211002, India.}
\maketitle
\begin{abstract}
The phenomenological consequences of unification of Einstein gravity and electromagnetism in an early phase of a Machian universe with a very small and uniform electrical charge density $\rho_q$  are explored. A form of the Strong Equivalence Principle for unified electrogravity is first formulated, and it immediately leads to (i) the empirical Schuster-Blackett law relating the magnetic moments and angular momenta of neutral astronomical bodies, (ii) an analogous relation between the linear acceleration of neutral massive bodies and associated electric fields, (iii) gravitational lensing in excess of Einstein gravity, and, with the additional assumption of scaling, to (iv) the Wesson relation between the angular momentum and the square of the mass of astronomical bodies.
Incorporation of Sciama's version of Mach's principle leads to a new post-Newtonian dynamics (in the weak field limit of gravity alone without electromagnetism) that predicts flat rotation curves of galaxies without the need of dark matter haloes. Finally, it is shown that the unified theory with a broken symmetry predicts a flat expanding universe with a cosmological term intimately related to electrogravity unification, and can explain WMAP data with a single free parameter. WMAP data require $\rho_q =6.1\times 10^{-43}$ C/cc which is too small to be detected at present.
\end{abstract}
\section{Introduction}

Gravity and Electromagnetism are the only two long range fundamental forces in the universe we know of, and their presence is ubiquitous. It was Einstein who first attempted a unified classical theory of these two forces, but such efforts were eventually given up \cite{uni}. One of the striking examples of a possible link between the two forces is the Schuster-Blackett law relating the angular momenta and magnetic moments of a remarkably large number and class of astronomical bodies \cite{sb, sirag}. Einstein had proposed a similar relationship in 1924 to account for terrestrial and solar magnetism \cite{eins} which is too large to understand in terms of charged convection currents inside the bodies. His motivation was to search for a unified field underlying gravity and electromagnetism.

The two main difficulties that stood in the way of a geometrical unification of the two forces were (a) the absence of a Strong Equivalence Principle for electrodynamics and (b) the enormous difference between their coupling strengths. The concept of a broken symmetry that is restored in some regime did not exist in those days.

The Equivalence Principle is one of the most fundamental principles in physics and, together with the requirement of covariance of the laws of physics under the most general coordinate transformations, constitutes the physical basis of Einstein's General Theory of Relativity (GR). Many variations of this principle, not all of them equivalent, exist in the literature, starting with Galileo and Newton down to Einstein and later authors such as Dicke \cite{dicke} who laid down the basis for experimental tests of the principle. Two of the key statements of the principle are:

{\flushleft {\em {\bf Universality of Free Fall or Weak Equivalence Principle} (WEP)}}
\vskip 0.1in
{\em All test bodies fall in a gravitational field with the same acceleration regardless of their mass or internal composition.}
\vskip 0.1in

This is the form associated with Galileo's law of falling bodies and Newton's theory of gravity which uses the equality of gravitational and inertial mass and was confirmed to high precision by the E\"{o}tv\"{o}s experiment. In this form it is usually referred to as the {\em Weak Equivalence Principle} (WEP). 

{\flushleft {\em {\bf Strong (or Einstein's) Equivalence Principle} (SEP)}}
\vskip 0.1in

{\em For every infinitesimally small world region in which space-time variations of gravity can be neglected, there always exists a coordinate system in which gravitation has no influence either on the motion of test particles or any other physical process.}   
\vskip 0.1in

This formulation \cite{pauli} can be summarized as: in an infinitesimal world region a homogeneous gravitational field can be transformed away. An observer in a freely falling lift, for example, cannot detect any gravitational effect on any object {\em inside the lift}---the homogeneous gravitational field {\em inside the lift} at rest on earth is transformed away inside the freely falling lift. This is the physical content of SEP. Unlike WEP, SEP is deeply connected with the geometry of General Relativity (GR) which treats space-time as a pseudo-Riemannian manifold with Lorentzian signature. The tangent plane to such a manifold is flat Minkowski space-time without gravity.

It must be pointed out, however, that for practical tests of the principle it is often difficult to specify the region $R$ over which the gravitational field is strictly uniform and can be transformed away, i.e. how small the `lift' should be in order, for example, to avoid tidal effects. The choice of $R$ does not depend on any reference frame. This is why some would prefer the following statement \cite{mathes}:
\vskip 0.1in

{\em For every infinitesimally small world region in which space-time variations of gravity can be neglected, there always exists a coordinate frame in which, from
the point of view of the co-moving observer, gravitation appears to have no influence on the motion of test particles, or on any other natural phenomenon,
but only as to local measurements that ignore the motion of test objects with respect to the source(s) of the field.}
\vskip 0.1in
 
It is generally believed that no analogous principle exists for electrodynamics. The reason is that unlike gravity which is always attractive, electromagnetism can be both attractive and repulsive. Furthermore, although the ratio of the gravitational and inertial mass is strictly unity for all matter, the ratio $q/m$ of electric charge to mass varies from particle to particle, violating WEP.

In order to probe a deeper level of unification of the forces behind all these apparent differences betweeen them, it is first necessary to seek a form of SEP that could be true in some symmetric phase of the universe in which gravity and electromagnetism were fully unified. Such a symmetry is clearly broken in the present universe, and hence a mechanism and measure of the symmetry breaking must also be found. In addition, in the spirit of Einstein, Mach's principle, as quantitatively enunciated by Sciama \cite{sciama}, will also be invoked. The consequence of putting these basic principles together is unexpectedly rewarding---a wide variety of astrophysical phenomena get correlated, like flat rotation curves of galaxies without dark matter, the Schuster-Blackett law, the Wesson relation between the angular momentum and the square of the mass of astronomical bodies such as planets all the way up to galactic clusters \cite{wesson}, and a flat universe with dark energy in the form of a cosmological constant. 

It must be emphasized that finding an alternative theory of these diverse phenomena outside current mainstream physics is not the motivating factor of this paper, and consequently it must be admitted that these results provide valuable insights into electrogravity symmetry and cosmology that may not be entirely fortuitous. {\em No claim, however, is being made that these well known classical principles alone constitute the only complete understanding of these varied phenomena}.  

In order to proceed further, it will be convenient first to display Einstein's field equations in a form similar to that of Maxwell's electrodynamics.

\section{Gravitomagnetism (GEM)}

As is well known \cite{wheeler}, in the limit of a weak gravitational field Einstein's field equations
\beq
R_{\mu\nu} - \frac{1}{2}g_{\mu\nu} R = -\frac{8\pi G}{c^4} T_{\mu\nu}\label{eins}
\eeq
reduce to the linear form
\beq
\Box \bar{h}_{\mu\nu} = -\frac{16\pi G}{c^4}T_{\mu\nu}.\label{1}
\eeq
This can be shown by writing the metric as $g_{\mu\nu} = \eta_{\mu\nu} + h_{\mu\nu}$ where $\eta_{\mu\nu} = (-1,1,1,1)$ is the flat Minkowski metric and $h_{\mu\nu} << 1$ is a small perturbation, and using the trace reversed form  $\bar{h}_{\mu\nu} = h_{\mu\nu} - \frac{1}{2}\eta_{\mu\nu}h$ with $h = \eta^{\mu\nu} h_{\mu\nu}$. The potentials $h_{\mu\nu}$ are gauge dependent, and hence the Lorentz gauge $\bar{h}_{\mu\nu\,,\nu}=0$ must also be used. The particular solution of this equation is the retarded function
\beq
\bar{h}_{\mu\nu}(\vec{x},t) = \frac{4G}{c^4} \int \frac{T_{\mu\nu}(ct - |\vec{x} - \vec{x}^{'}|) }{|\vec{x} - \vec{x}^{'}|} d^3 \vec{x}^{'}
\eeq
Defining the mass density $\rho_m=T^{00}/c^2$, the mass 4-current desnsity $j_m^i = T^{0i}/c$, and $A^0_g = \Phi_g$ where $\Phi_g$ is the Newtonian potential, and using the approximations $|T^{00}| >> |T^{0i}| >> |T^{ij}|$ and ignoring all terms of order $1/c^4$ and smaller, one can write the line element \cite{mashhoon1} 
\ben
ds^2 &=& - (1 + \frac{2\Phi_g}{c^2})\,c^2 d t^2  + \frac{4}{c} A^g_i dx_i dt + (1 - \frac{2\Phi_g}{c^2})\delta_{i j}dx_i dx_j, \label{met1}
\een
This is the line element due to a slowly rotating spherical body.

In analogy with electrodynamics, one can define the fields ${\vec E}_g$ and ${\vec B}_g$ by 
\ben 
{\vec E}_g &=& -\vec{\nabla} \Phi_g - \frac{1}{c}\frac{\partial}{\partial t} \left(\frac{1}{2}{\vec
A}_g \right),\\
{\vec B}_g &=& \vec{\nabla} \times {\vec A}_g.
\een 
These are called respectively the gravitoelectric and gravitomagnetic fields. One can also write the covariant form of the field equations
\ben
2 F_g^{\mu\nu} &=& \partial^\mu A^\nu_g - \partial^\nu A^\mu_g,\\
\partial_\mu F_g^{\mu\nu} &=& - 4\pi G j^\nu,\label{minus}
\een
where $A_\mu^g = (-\Phi_g = -A^g_0, {\vec A}^g)$. In spite of these formal similarities, there are fundamental differences with electrodynamics reflected in the minus sign in (\ref{minus}) because gravity is always attractive, and factors of 2 because gravity is spin 2. The principal difference with Newtonian gravity is the existence of the gravitomagnetic field ${\vec B}_g$.

The gravitational analog of Lorentz's ponderomotive force density \cite{note} can be defined as
\beq
f^g_\mu = F^g_{\mu\nu}j^\nu_g 
\eeq 
with
\ben
{\vec f}^g &=& \rho \left[{\vec E}_g + \frac{2}{c} {\vec v} \times {\vec B}_g\right],\\
f^g_0 &=& \rho_m \frac{{\vec E}_g. {\vec v}}{c},
\een
where ${\vec f}_g$ is the analog of the Lorentz force density. The equation of motion of a test particle of rest mass $m_0$
is then
\beq
\vec{F}^g = m_0 {\vec a} =\gamma m_0 \left[{\vec E}_g + \frac{2}{c} {\vec v} \times {\vec B}_g\right],\label{LF1}
\eeq
and $F^g_0 = (m_0/c) \vec{v}.\vec{E}_g$. 
The mass $m_0$ drops out of this equation and WEP holds, as expected. Clearly, a coordinate transformation to a frame with acceleration ${\vec a}$ transforms away the gravitational field, and SEP also holds. In such an accelerated frame 
\beq
{\vec E}_g + \frac{2}{c} {\vec v} \times {\vec B}_g = 0,\label{gem}
\eeq
hence $F^g_0 =0$, and therefore $F^g_\mu = 0$. Thus, all gravitational effects are transformed away.
The weak field approximation involved in the GEM equations is sufficient for the elucidation of the equivalence principle which is a statement about the neighbourhoods of the flat tangent planes to the pseuo-Riemannian manifold. 

\section{Electrogravity and the Strong Equivalence Principle}

Having displayed the formal similarity (but no isomorphism) between classical General Relativistic gravity and classical electromagnetism, we will now move on to consider electrodynamics itself. Let us start with Lorentz's ponderomotive force 
\beq
\phi_\mu = F_{\mu\nu}j^\nu \label{pond}
\eeq
where $F_{\mu\nu}$ is the electromagnetic field tensor and $j^\nu = ((\rho_q)_0/c) v^\nu$ is the conserved charge current-density 4-vector, $(\rho_q)_0$ being the charge density in the rest frame of the charges. The space component of $\phi_\mu$ is the Lorentz force density
\beq
{\vec \phi} = \rho_q \left[{\vec E} + \frac{1}{c} {\vec v} \times {\vec B}\right],
\eeq
and the time component $\phi_0 = \rho_q ({\vec E}. {\vec v}/c)$ is the power per unit volume divided by $c$. It follows from (\ref{pond}) that $\phi_\mu j^\mu = 0$. 

Consider now a homogeneous electromagnetic field in an infinitesimal world region $R$ sufficiently distant from its source bodies, and a test charge density $\rho_q$ in $R$ of negligible rest mass density $\rho_{m}$ so that all gravitational effects can be ignored. The equation of motion of the test body is
\beq
\vec{\phi}= \rho_{m} \vec{a} =  \rho_q \left[{\vec E} + \frac{1}{c} {\vec v} \times {\vec B}\right],\label{eqn}
\eeq
and $\phi_0 = (\rho_q/c) \vec{v}.\vec{E}$. In this case, the acceleration $\vec{a}$ is dependent on the ratio $\rho_q/\rho_m$, and so WEP does not hold. Nevertheless, since $\rho_q/\rho_m \neq 0$, to an observer at rest in a coordinate frame with acceleration $\vec{a}$ the condition
\beq
{\vec E} + \frac{1}{c} {\vec v} \times {\vec B} = 0
\eeq
must hold. This would imply that $\phi_0 =0$ and hence $\phi_\mu =0$ in that frame. In other words, an observer at rest in that frame will be unable to detect any electromagnetic influence on the test charge (barring its self-field which is to be ignored). Hence, transformation to such an accelerated coordinate frame `transforms away' the electromagnetic field, leaving the charged body in a state of rest.

Nevertheless, there {\em is} a fundamental difference with gravity, namely that the choice of the accelerated frame depends on the factor $\rho_q/\rho_m$ which is not universal, violating WEP. This appears to stand in the way of a true unification of the two forces in a geometric theory. There {\em is}, however, a way of overcoming this problem, and that is to invoke Mach's principle to determine the property of inertia of charged bodies by the large scale distribution of mass and charge in the universe. 

\section{Mach's Principle and the Origin of Inertia}
Let us consider a universe of mass density $\rho_m(\tau)$ which is uniform on a large scale at all proper times $\tau$. Then the gravitational scalar potential at any test body is
\beq
\Phi_g =  G \int_\sigma \frac{\rho_m(\tau)}{r}\eta^\mu d\sigma_\mu = - 2\pi G\rho_m(\tau) c^2\tau^2
\eeq
where $\sigma$ is a space-like surface, $\eta^\mu$ is a unit space-like four-vector (i.e. $\eta^\mu \eta_\mu = -1$) and the radius of the universe $R(\tau)$ has been set equal to $c\tau$. Henceforth we will write $\rho_m(\tau)=\rho_m$. There is no vector potential $\vec{A}_g$ in this case.

Let us now calculate the potentials when a test body moves with a small velocity $\vec{v}(t)$ relative to the uniform universe, which is equivalent to the universe moving with the velocity $-\vec{v}(t)$ relative to the test body at rest. In addition to the Hubble effect due to the expansion of the universe, there will also be a Doppler shift observable at the position of the body at time $t$ corresponding to $\vec{v}(t)$ from all parts of the universe.
Hence, the velocity observable at the position of the body at time $t$ must be taken to be $\vec{v}(t) + \vec{r}/\tau$ where $\tau$ is the inverse of the Hubble parameter $H$. Ignoring terms of the order $v/c$ and higher, we again have
\beq
\Phi_g = - 2\pi G\rho_m c^2\tau^2\label{phi}
\eeq
There is now, however, a vector potential
\beq
\vec{A}_g = -G \int_V \frac{\vec{v}\rho_m}{c r} dV = \frac{\Phi_g}{c}\vec{v}(t). 
\eeq
Hence, the gravitoelectric field created is
\beq
\vec{E}_g = - \vec{\nabla}\Phi_g - \frac{1}{c}\frac{\partial \vec{A}_g}{\partial t} = -\frac{\Phi_g}{c^2}\frac{\partial \vec{v}}{\partial t}
\eeq
because by assumption $\vec{\nabla}\Phi_g = 0$. The gravitomagnetic field is
\beq
\vec{B}_g = {\rm curl} \vec{A}_g.
\eeq
Let there be a body of gravitational mass $M$ at a distance $r < c\tau$ from the test body in the otherwise uniform universe. Then in the rest frame of the test body the local field of the body is
\beq
-\frac{GM}{r^2} \hat{r}
\eeq
Let us now invoke the Sciama postulate \cite{sciama} to incorporate Mach's principle : 
\vskip 0.1in
{\em In the rest-frame of any body the total gravitational field at the body arising from all the other matter in the universe is zero.}
\vskip 0.1in
This ensures that the origin of inertia of an apparently isolated body via Mach's principle is consistent with the {\em apparent irrelevance} of the properties of the universe as a whole. 
Then, it follows (from Gravitoelectrodynamics) that
\beq 
-\frac{GM}{r^2} \hat{r} + \left[{\vec E}_g + \frac{2}{c} {\vec v} \times {\vec B}_g\right] = 0.
\eeq 
On taking the scalar product of both sides with $\hat{r}$, one gets
\beq
-\frac{GM}{r^2} = \frac{\Phi_g}{c^2} \left(\frac{dv}{dt} - 2v \frac{\partial v}{\partial r} +2\frac{v^2}{r}\right). \label{X} 
\eeq
This shows that once Mach's principle is taken into account, one cannot drop the gravitomagnetic term compared to the gravitoelectric term in the gravito-Lorentz force, as is usually done because of the $v/c$ factor.
The overall factor $\Phi_g/c^2$ determines the inertial property of the body due to the rest of the matter in the universe. In order to have Newton's law as a limiting case, one must choose $\Phi_g/c^2 = -1$ to get 
\beq
a_g = \hat{r}.\left[{\vec E}_g + \frac{2}{c} {\vec v} \times {\vec B}_g\right] =\left(\frac{dv}{dt} - 2v \frac{\partial v}{\partial r}+2\frac{v^2}{r}\right) = \frac{GM}{r^2}. \label{Y}  
\eeq
This expression reduces to Newton's law in the absence of the second and third terms which are due to the gravitomagnetic field. {\em It shows that, under certain conditions, Newton's laws of motion can be very accurate despite their apparent complete lack of reference to the properties of the universe} \cite{unni1}.

One can have a parallel derivation of Coulomb's law by extending Sciama's criterion to electrodynamics:
\vskip 0.1in
{\em In the rest-frame of any charged body the total electromagnetic field at the body arising from all the other charges in the universe is zero}. 
\vskip 0.1in
In a universe of uniform charge density on a large scale, the equation analogous to (\ref{X}) is  
\beq
\pm\frac{Q}{4\pi\epsilon_0 r^2} = \frac{\Phi_q}{c^2}\frac{m}{q} a_q \label{X2} 
\eeq
if $\Phi_q/c^2 = -1$, $\Phi_q$ here being the electrostatic potential induced by charges. Hence, $\Phi_q=\Phi_g$. This results in 
\beq
a_q =\pm \frac{q}{m}\hat{r}. \left[{\vec E} + \frac{1}{c} {\vec v} \times {\vec B}\right] = \pm \frac{q}{m}\frac{Q}{4\pi\epsilon_0 r^2},\label{Y2} 
\eeq
which is indeed Coulomb's law when $a_q$ is rectilinear.  

It is clear from (\ref{Y}) and (\ref{Y2}) that the net acceleration of a test mass $m$ of charge $q$ would be
\ben
a_g - a_q &=&\frac{GM}{r^2}\left(1 \mp \frac{1}{4\pi G\epsilon_0}\frac{Qq}{Mm}\right)\nonumber\\
&=& 0, \,\,2\frac{GM}{r^2}
\een
if 
\beq
\frac{Mm}{Qq} = \frac{1}{4\pi G\epsilon_0}.
\eeq
If one further postulates that $ m/q=M/Q=\zeta = 1/\sqrt{4\pi\epsilon_0 G}$ {\em for all matter on a large scale} (i.e. there is no electrically neutral matter on a large scale) such that the universe has a charge density (positive or negative) $\rho_q = \zeta^{-1}\rho_{qm}$, the net acceleration of a test body in such a universe would be
\ben
\vec{a} &=&\left[({\vec E}_g - \zeta^{-1}{\vec E}) + \frac{1}{c} {\vec v} \times (2{\vec B}_g - \zeta^{-1}{\vec B})\right] =0,\label{elec}
\een
the repulsive acceleration due to electromagnetism balancing the attractive acceleration due to gravity.
The Universality of Free Fall (WEP) for all matter would clearly hold on a large scale in such a universe {\em even when the balance between electromagnetism and gravity is lost, the net non-zero $\vec{a}$ being universal for all matter}. Hence, a Strong Equivalence Principle would also hold on a large scale in such a universe, and we are led to the following statement:

{\flushleft {\em {\bf Strong Equivalence Principle in a Unified Theory}}}
\vskip 0.1in
{\em In a universe with a uniform mass and charge density and $\zeta = \rho_{qm}/\rho_q = 1/\sqrt{4\pi\epsilon_0 G}$, there always exists a coordinate system in which gravity and electromagnetism have no influence either on the motion of bodies or any other physical process.}   
\vskip 0.1in

The right-hand side of Einstein's field equations in such a unified universe would have two terms, namely 
\ben
-\frac{8\pi G}{c^4} T^r_{\mu\nu} - \frac{2}{\epsilon_0 c^4} T^{q}_{\mu\nu},\nonumber\\\epsilon_0 =\frac{1}{4\pi G} \,\,{\rm or}\,\, \zeta =1 {\rm kg/C}.
\een
The first term represents the contribution of radiation density $\rho_r = T^r_{00}/c^2$ and the second term represents that of a charged fluid of density $\rho_q= T^q_{00}/c^2$. The following relations are assumed to hold in such a universe:
\ben
\rho_r &=& \rho_{qm} = \zeta \rho_q,\\
\Phi_q &=& \Phi_g,\\
A^q_i &=& A^g_i.  
\een 
Recalling the result (\ref{met1}) in the weak field approximation, it is clear that the line element of such a universe will be
\ben
ds^2 &=& - (1 + \frac{2(\Phi_g+\Phi_q)}{ c^2})\,c^2 d t^2  + \frac{8}{c} A^g_i dx_i dt + (1 - \frac{2(\Phi_g +\Phi_q)}{ c^2})\delta_{i j}dx_i dx_j\label{met3a}\\
&=&- (1 + \frac{4\Phi_g}{ c^2})\,c^2 d t^2  + \frac{8}{c} A^g_i dx_i dt + (1 - \frac{4\Phi_g}{ c^2})\delta_{i j}dx_i dx_j,\label{met3}
\een
and hence {\em there is an exact doubling of the gravitational potentials} in this symmetric phase. As a result, {\em the theory predicts twice the gravitational lensing predicted by Einstein gravity alone in such a case}.

The breaking of symmetry would result when there is neutral matter and the total density $\rho_m$ due to radiation and matter is different from $\rho_{qm}$, as we will see in more detail later. Having postulated a universe with electrogravity symmetry in which the ratio $m/q =\zeta$ is a universal constant, let us now see what observable predictions follow from it. The physical content of the postulate can only be judged by the extent to which its predictions are in agreement with observations. 

{\flushleft {\em Cosmological Implications: The Schuster-Blackett Law}}
\vskip 0.1in

Since the same acceleration transforms away both gravitational and electromagnetic fields locally (SEP in the unified universe), it follows from Eqn. (\ref{elec}) that 
\beq
\vec{E} = \zeta \vec{E}_g,\,\,\,\,\,\,\vec{B} = 2\zeta \vec{B}_g. 
\eeq
Since a rotating massive object produces a gravitomagnetic field, this implies that such an object must also produce a magnetic field. Hence, the ratio of the magnetic moment $\mu$ and the angular momentum $J$ of a rotating body must be
\beq
\frac{\mu}{J} = \frac{1}{2\zeta} = \frac{\sqrt{4\pi G\epsilon_0}}{2} = \frac{\sqrt{G}}{2\sqrt{k}}, 
\eeq
where $k = 1/4\pi\epsilon_0$ is the Coulomb constant in SI units. The numerical value of this ratio is $4.3 \times 10^{-11}\,{\rm C/kg}$ in the present universe. This is indeed the empirical Schuster-Blackett law mentioned earlier within an overall form factor $\beta$ of order unity. It is well known that though cosmic magnetic fields pervade the universe, there is no accepted theory of their origin. Nonminimal gravitational-electromagnetic coupling (NMGEC) has been suggested by some to be the source of the Schuster-Blackett law \cite{wood} and of the intense magnetic fields near rotating black holes, connected with quasars and gamma-ray bursts. \cite{oph1, oph3, oph2}. It is clear from our considerations that the Schuster-Blackett law is a simple and direct consequence of the Strong Equivalence Principle in a unified theory. There is therefore considerable empirical support in favour of a Unified Theory of gravitation and electromagnetism. Since $\zeta =1$ in the fully symmetric limit, the Schuster-Blackett law reflects a badly broken symmetry.

A similar relation between the magnitudes of the electric fields $\vert \vec{E}\vert$ and linear accelerations $\vert \vec{a}\vert$ of neutral mass densities, namely
\beq
\frac{\vert\vec{E}\vert}{\vert\vec{a}\vert} =\zeta,\label{Ea}
\eeq 
must also hold in the universe, and is a definite prediction of SEP and electrogravity symmetry. Colliding galaxy clusters with highly accelerated x-ray emitting jets are possible testing grounds for this law. 

Yet another prediction follows from the relation $q^2/m^2 = 4\pi\epsilon_0 G$ if one makes an additional assumption. Let $e$ and $m$ be the charge and mass of a fundamental fermion. Then $e^2/m^2 = 4\pi\epsilon_0 \hbar c \alpha/m^2= 4\pi\epsilon_0 G$, where $\alpha$ is the fine structure constant. Therefore, $\hbar/m^2 = 2\sigma/m^2 = G/\alpha c$ where $\sigma = \hbar/2$ is the spin of the particle. If one assumes that $\sigma$ amd $m^2$ scale by the same factor, one is led to the prediction
\beq
\frac{J}{M^2} = \frac{G}{2\alpha c} \simeq 1.52\times 10^{-16}\,{\rm cm^2 g^{-1}s^{-1}}
\eeq
for macroscopic objects of angular momentum $J$ and mass $M$. This relation is well established for astronomical objects of various types from asteroids to the Local Supercluster \cite{wesson}. The above simple derivation of the relation is indicative of its origin in the Strong Equivalence Principle and an electrogravity unified universe with a broken symmetry.

To explore further the consequences of electrogravity symmetry and its breaking, let us first see how Mach's principle can be incorporated into the theory.

\section{Cosmological Implications of Mach's Principle}
The second term term in Eqn. (\ref{Y}) is a centrifugal acceleration and is essential for understanding the dynamics of non-circular motion in a plane, such as in spiral galaxies. On integration it gives
\beq
v^2 = GM\left(\frac{1}{r} -\frac{1}{r_0}\right)  + v_0^2 = \frac{GM}{r} + \left( v_0^2 -\frac{GM}{r_0}\right) \label{Q} 
\eeq
where $v_0$ is the velocity at a distance $r_0$ from the origin. The first term on the right hand side is the Newtonian term, and it falls off asymptotically as $r^{-1}$. The second term is a constant that vanishes only for a single value of $v_0$, namely $|v_0| = \sqrt{GM/r_0}$, and immediately {\em predicts flat rotation curves} for spiral galaxies for all other values of $|v_0| > \sqrt{GM/r_0}$ as $r\rightarrow \infty$ {\em without requiring any dark matter}. Writing the second term as $\beta v_c^2$ where $\beta$ and $v_c$ are arbitrary parameters in place of $(v_0, r_0)$, equation (\ref{Q}) can be written as
\beq
g = \frac{v^2}{r} = \frac{GM}{r^2} + \beta\frac{v_c^2}{r} = g_N + \beta\frac{v_c^2}{r}, 
\eeq
and hence 
\beq
g_N = g\left(1 - \beta\frac{v_c^2}{gr}\right) \equiv g\mu.
\eeq
This is a modification of the Newtonian acceleration that is a direct consequence of the Mach principle plus gravitoelectrodynamics, a weak field limit of General Relativity. In regions where $g$ is small (e.g. very large $r$), the effective gravitational acceleration becomes $\simeq \beta v_c^2/r$. The Newtonian limit corresponds to $\beta =0$.

This is reminiscent of modified Newtonian dynamics (MOND) \cite{milgrom}. Indeed, if one sets $v_c^2 = \sqrt{GM a_0}$ where $a_0$ is an arbitrary parameter of the dimension of acceleration and assumes that $\beta = \beta(x)$ where $x= g/a_0$, one can write
\ben
g_N &=& \mu (x) g,\label{mond}\\
M &=& \frac{v_c^4}{Ga_0}, \label{TF}
\een
with $\mu =1$ when $x\gg 1$.
These are the basic equations of modified Newtonian dynamics. With $\mu = x/\sqrt{1 + x^2},\,x \lesssim 1$ and $\mu = x/(1 + x),\,\,x\gtrsim 10$, $a_0 \simeq 1.2\times 10^{-8} {\rm cm/s^2}$, one gets good fits to astrometric data covering planetary motion in the solar system \cite{sereno} and galaxies \cite{mond} as well as the Tully-Fisher relation \cite{TF,mac} between the intrinsic luminosities $L_H$ (proportional to the baryonic masses $M$) of such galaxies and $v^4_c$ (Eqn. (\ref{TF})) without requiring any significant contribution from non-baryonic ``dark matter''. 

The most distinguishing feature of the Bullet and similar clusters is the separation of the hot x-ray emitting gases from the two centres of gravitational lensing where the putative dark matter is supposed to be concentrated. This separation and resultant gravitational lensing pattern can be qualitatively understood as the result of the action of the electric fields created by the linear accelerations of the two neutral gas clouds in opposite directions (in accordance with Eqn. (\ref{Ea})) on the background uniform charge distribution. The electric fields accelerate the charges in opposite directions with acceleration $\vert\vec{a}_q\vert = \rho_{qm}^{-1}\rho_q \vert\vec{E}\vert =\zeta^{-1} \zeta f^{'}\vert\vec{a}\vert =f^{'}\vert\vec{a}\vert$. Hence, if $f^{'}>1$, the charges overtake the gas clouds and build up in front of them, forming a dipole configuration. These charge accumulations, like dark matter, act as gravitational lenses in accordance with the result (\ref{met3a}) of electrogravity symmetry. Whether or not this is the case can, of course, only be verified by numerical simulations which have not been carried out yet.

\section{Other Cosmological Implications}
Finally, let us consider the difference of the gravitational and electromagnetic potentials when the Hubble parameter is $H$, namely
\ben
\phi &=& - 2\pi G(\rho_m - \rho_{qm}) c^2/H^2\label{phi2}
\een
where $\rho_m c^2$ is the energy density due to the uniform distribution of neutral mass and radiation and $\rho_{qm} c^2$ that due to uniform charged mass distribution in the universe. 
Thus,
\beq
H^2 = \left(\frac{\dot{a}}{a}\right)^2 = \frac{8\pi G (\rho_m - \rho_{qm})}{3}.\left(-\frac{3 c^2}{4\phi}\right)= \left(\frac{8\pi G \rho_m}{3} - \frac{\Lambda c^2}{3}\right)\chi \label{H} 
\eeq
where $\Lambda c^2= 8\pi G \rho_{qm}$ and $\chi = -3c^2/4\phi$. This reduces to the standard Friedmann equation with $k=0$ (a flat universe) and a negative cosmological constant when $\chi = 1$. It follows therefore that for $\chi \neq 1$
\beq
\dot{H} = \frac{4\pi G}{3H}\frac{d}{d t}(\rho_m \chi)\label{dotH}
\eeq
and
\ben
\frac{\ddot{a}}{a} &=& H^2 + \dot{H} =  \frac{8\pi G}{3}\left(\rho_m \chi + \frac{1}{2H} \frac{d}{dt}({\rho}_m \chi) \right) - \frac{\Lambda c^2}{3}\chi\nonumber\\
&=& - \frac{4\pi G}{3}\left(\rho_m + \frac{3p}{c^2}\right)\chi - \frac{\Lambda c^2}{3}\chi\label{doubledot}
\een
provided 
\beq
\frac{d}{dt}({\rho}_m \chi) = -3 H(1+w) \rho_m \chi,\label{dot}
\eeq
where $w = p/\rho_m c^2$. Hence
\beq
\frac{\rho_m \chi}{(\rho_m \chi)_0} = {\rm exp}\left(-3\int_{t_0}^t H(1+w) dt\right) = {\rm exp}\left(-3\int_{a_0}^a \frac{da}{a}(1+w)\right)  = \left(\frac{a}{a_0}\right)^{-3(1+w)}\label{rhofall}
\eeq
When $p = 0$, $w=0$ and
\beq
\frac{\rho_m \chi}{(\rho_m \chi)_0} =  \left(\frac{a}{a_0}\right)^{-3}.
\eeq
If $\rho_q \propto a^{-3(1+w)}$, $\zeta$ must evolve like $a^{+3(1+w)}$ so that $\rho_{qm}=\zeta\rho_q$ is a constant with equation of state $w=-1$, and acts like a cosmological constant. Since like charges repel, $\rho_q$ should have negative pressure. Let $w_q = -1 + 2\delta/3$ where $\delta$ is a small parameter. Then $\zeta \propto a^{2\delta}$ and $G\epsilon_0 \propto a^{-\delta}$.  

Finally, it follows from (\ref{phi2}) that at present
$$\rho_m -\rho_{qm}  \simeq -\frac{\phi/c^2}{2\pi G \tau_0^2}$$
Using $G = 6.67 \times 10^{-8} {\rm cm^3 g^{-1} s^{-2}}$, $\epsilon_0 = 8.85 \times 10^{-21}{\rm C^2 cm^{-3}g^{-1} s^2} $ and $\tau_0 \simeq 4 \times 10^{17}$ sec (WMAP)\cite{wmap}, one can estimate
\beq
\rho_m - \rho_{qm} \simeq -1.5\times 10^{-29}\frac{\phi}{c^2} {\rm g/cc}. 
\eeq 
Since $k=0$ and the universe is flat,
\beq
\rho_{tot} = \rho_m + \rho_{qm} = \rho_c = 10^{-29}\,{\rm g/cc}.
\eeq
WMAP data have confirmed that the universe is indeed flat, and $71\%$ of the observed density has been ascribed to dark energy and the rest to baryonic and dark matter. Hence a good fit to the WMAP data is obtained if one sets $\phi/c^2 \simeq 0.28$ (i.e. $\chi =- 0.21$). Then
\ben
\rho_m &\simeq & 0.29 \times 10^{-29}\,{\rm g/cc},\\
\rho_{qm} &\simeq & 0.71 \times 10^{-29}\,{\rm g/cc}.
\een
This gives a measure of the current breakdown of electrogravity symmetry. It also leads to the prediction 
\beq
\rho_q = \zeta^{-1} \rho_{qm}= \sqrt{4\pi G\epsilon_0}\, \rho_{qm} \simeq 6.1 \times 10^{-43}{\rm C/cc}.
\eeq 
Since the electron charge is $e = 1.60217657 \times 10^{-19}$ C, this is equivalent to having excess charge of approximately $3.8\times 10^{-24}$e per cc which will be nealy impossible to rule out except by making specific models in cosmological dynamics \cite{unni}. Such a miniscule uniform charge density $\rho_q$ is sufficient to generate the observed accelerated expansion of the universe because of the parameter $\zeta$ which is large. 
{\em The phenomenon of `dark energy' can thus be explained in terms of an extremely small charge density of the universe required for electrogravity symmetry.}

\section{Summary}
I have argued that a geometric unification of classical electrodynamics and gravity requires a Strong Equivalence Principle to hold in an electrogravity symmetric universe with a tiny electric charge density $\rho_q \simeq 10^{-43}$ C/cc and a universal mass to charge ratio $\zeta = m/q = \rho_{qm}/\rho_q=1/\sqrt{G\epsilon_0}$ over large scales. The fully symmetric limit corresponds to $\zeta = 1$ kg/C. In the present universe $\zeta$ is very large, indicating that the symmetry is badly broken. That immediately leads to the empirical Schuster-Blackett law and hence an explanation of the origin of ubiquitous astrophysical magnetic fields. In addition, it also leads to an analogous new prediction for the linear accelerations of neutral bodies and associated electric fields. With the additional assumption of scaling, it leads to the Wesson relation $J/M^2 = p$, a constant. In the fully symmetric limit, it predicts a doubling of the gravitational potentials. Mach's principle and Gravitoelectrodynamics, a weak field limit of General Relativity, automatically generate a post-Newtonian dynamics (POND) that predicts flat rotation curves of spiral galaxies without the need of dark matter haloes. With a suitable choice of parameters, one can derive the MOND relations (\ref{mond}, \ref{TF}) which give good fits to astrometric data. Mach's principle together with broken electrogravity symmetry leads to the prediction of a flat universe, and, given the values of the fundamental constants $G, \epsilon_0$, the Hubble parameter and a single free parameter $\phi/c^2$, to other cosmological predictions in good agreement with the WMAP data.  Neutral matter in the universe behaves like a pressure-less ($w=0$) ideal gas while the energy density $\rho_{qm}c^2$ due to the excess charge, required for electrogravity symmetry, has an equation of state $w = -1$ and acts like a cosmological constant that drives an accelerated expansion of the universe in the present epoch. It can account for $\sim 71\%$ of the total energy of the universe without requiring any other source of `dark energy'. 

Mach's principle ensures that the large scale physics of the cosmos is compatible with a Newton-Maxwellian local universe provided $\Phi_g/c^2 = \Phi_q/c^2 = -1$.

It must be emphasized that unlike the Lyttleton-Bondi cosmology \cite{bondi}, the total charge is assumed to be a constant, the excess charge density $\rho_q$ varies with the scale factor $a$, and there is no spontaneous creation of charge in the cosmology. This is possible if the universal parameter $G\epsilon_0$ varies over cosmological times in a predicted manner ($\propto a^{-\delta}$). If that is not observed, it would leave open the possibility of a Lyttleton-Bondi cosmology.

It is quite remarkable that just the combination of electrogravity symmetry and Mach's principle, which constitute what Einstein called a `principles theory' as opposed to a `constructive theory' to explain a particular type of phenomenon \cite{stach}, contains such a wealth of detailed information of the cosmos. The account given above is admittedly incomplete, at least in the sense that it is silent about the beginning, if any, of the universe and its ultimate fate. These are issues I wish to address in the future. 
  
As we will see in the next paper, it is possible to build a unified theory of gravity and electromagnetism by postulating a primordial affine manifold with non-symmetric connection which is `projective invariant'. This invariance is broken by the matter term in the Lagrangian, and the manifold splits into a symmetric part describing Einstein gravity and an antisymmetric part descrbing electromagnetism. 

\section{Acknowledgement}
I thank Mira Dey, Jishnu Dey, Saikat Chatterjee, Tarun Souradeep, Sanjit Mitra and C. S. Ummikrishnan for many helpful discussions. I also thank David Mathes and Paul Zelinsky for commenting on SEP in the first version of this paper, and Maldehai Milgrom for pointing out a slight defect in the discussion on MOND, which has now been corrected. Finally, I thank
the National Academy of Sciences, India for the grant of a Senior Scientist Platinum Jubilee Fellowship that enabled this work to be undertaken.

\end{document}